# FAST CALCULATION METHOD OF AVERAGE G-FACTOR FOR WAVE-CAIPI IMAGING

*Haifeng Wang[1], Zhilang Qiu[1, 2], Shi Su[1], Leslie Ying[3], Dong Liang[1]*

[1]Paul C. Lauterbur Research Center for Biomedical Imaging, Shenzhen Institutes of Advanced Technology, Chinese Academy of Sciences, Shenzhen, Guangdong 518055, China
[2]Shenzhen College of Advanced Technology, University of Chinese Academy of Sciences, Shenzhen, Guangdong 518055, China
[3]Department of Biomedical Engineering and Department of Electrical Engineering, University at Buffalo, The State University of New York, Buffalo, New York 14260, USA

**ABSTRACT**

Wave-CAIPI MR imaging is a 3D imaging technique which can uniformize the g-factor maps and significantly reduce g-factor penalty at high acceleration factors. But it is time-consuming to calculate the average g-factor penalty for optimizing the parameters of Wave-CAIPI. In this paper, we propose a novel fast calculation method to calculate the average g-factor in Wave-CAIPI imaging. Wherein, the g-factor value in the arbitrary (e.g. the central) position is separately calculated and then approximated to the average g-factor using Taylor linear approximation. The verification experiments have demonstrated that the average g-factors of Wave-CAIPI imaging which are calculated by the proposed method is consistent with the previous time-consuming theoretical calculation method and the conventional pseudo multiple replica method. Comparison experiments show that the proposed method is averagely about 1000 times faster than the previous theoretical calculation method and about 1700 times faster than the conventional pseudo multiple replica method.

## 1. INTRODUCTION

Parallel imaging techniques (pMRI) have been successfully applied to reduce the magnetic resonance scan time by acquiring undersampled k-space data [1, 2]. The SNR of pMRI is reduced by the square root of the acceleration factor due to k-space undersampling and the so-called g-factor depending on the encoding capabilities of the phased array coils [1]. Many methods have been proposed to reduce the g-factor penalty in parallel imaging [3-11]. A representative solution is the application of CAIPIRINHA concept, where the g-factor penalty is decreased and uniformed by controlling the aliasing pattern to make more effective use of coil sensitivity variations [3-6]. In 3D imaging, CAIPIRINHA is applied by modifying the gradient encoding scheme to shift the aliased slices (2D CAIPIRINHA) [5]. All of these CAIPIRINHA methods account for shifted aliasing pattern and make more effective use of the coil sensitivity variations in phase encoding and slice (or partition encoding) directions. However, the sensitivity variation in readout direction can also be utilized. One way is applying the CAIPIRINHA concept into non-Cartesian trajectories, such as radial CAIPIRINHA where the RF phase cycling is performed across neighboring radial spokes [6], and in spiral trajectory with application of z-encoding blips during the spiral readout [7]. Another way is modifying the Cartesian trajectories to non-Cartesian trajectories by imparting further phase modulations during the readouts, such as Zig-Zag CAIPIRINHA [8], bunched phase encoding (BPE) [9], Wave-CAIPI [10-12], etc. The recent Wave-CAIPI technique combines 2D CAIPIRINHA [5] and BPE to fully take advantages of coil sensitivity variations in three directions and significantly reduce g-factor penalty for accelerated 3D imaging [10]. It is a data acquisition strategy which is implemented by additional wave gradients simultaneously during the readout along with 2D CAIPIRINHA sampling scheme. The Wave-CAIPI technique has been extended to simultaneous multi-slice imaging (SMS) by combining blipped CAIPIRINHA [3] and BPE for accelerated SMS imaging [11].

To optimize the sampling patterns in parallel imaging, some methods have been explored, wherein accurate evaluation and computationally efficient methods were needed [13-15]. Although some evaluation criteria based on spectral moment have been explored [15], the g-factor related criteria were the most commonly used. For example, in 2D CAIPIRINHA, the average and the maximum g-factor have been used to find the optimal sampling pattern [13, 14]. In Wave-CAIPI imaging, the average and maximum g-factor have also be used as metric to experimentally evaluate the sampling pattern and wave gradient parameters [16, 17]. Although both of the theoretical calculation [11] and pseudo multiple replica methods [18] are able to estimate the average g-factor, these conventional methods are time-consuming and unpractical for the optimization of

This work is supported in part by National Natural Science Foundation of China (61871373, 81729003) and Natural Science Foundation of Guangdong Province (2018A0303130132).

sampling pattern and wave gradient parameters. Since the average g-factor is often used as a metric for selecting optimal sampling patterns or parameters, its fast calculation should be very useful in Wave-CAIPI imaging.

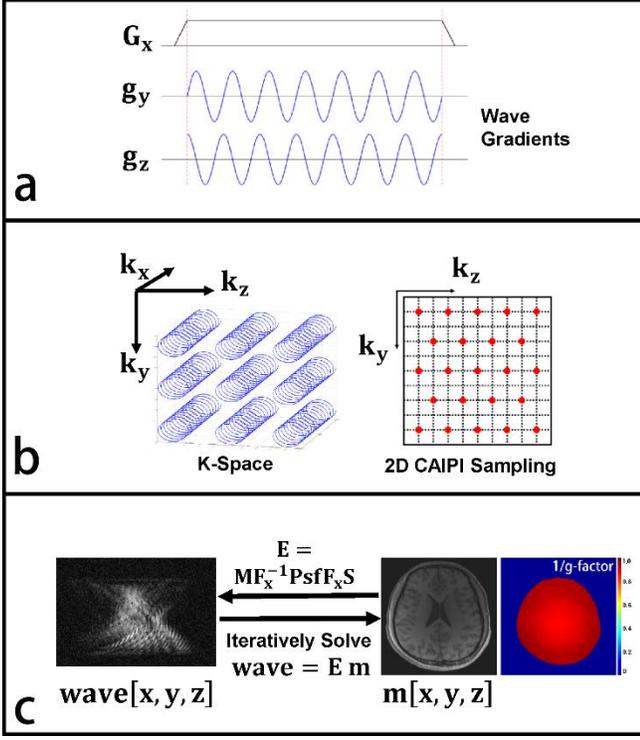

Fig. 1. The schematic diagram of Wave-CAIPI imaging. Additional wave gradients are applied to generate corkscrew k-space trajectories (a). Meanwhile, the 2D CAIPIRINHA sampling scheme is employed (b). The image encoding and reconstruction are as the generalized SENSE model (c).

In this paper, we propose a novel fast calculation method to calculate the average g-factor in Wave-CAIPI imaging, which is much faster than the conventional other two methods, the theoretical calculation and pseudo multiple replica. At first, the g-factor penalty in the arbitrary (e.g. the central) position is separately calculated, and then approximated to the average g-factor penalty using Taylor linear approximation [19]. The experiments demonstrated that the proposed fast calculation method was averagely approximately 1000 times faster than the conventional theoretical calculation method and about 1700 times faster than the pseudo multiple replica method. The preliminary results have been published previously in the Ref. [20].

## 2. THEORY
### 2.1. Overview of Wave-CAIPI
Wave-CAIPI is a 3D imaging technique to significantly reduce g-factor penalty at high acceleration factors. As seen as Fig. 1(a) and (b), the acquisition strategy is implemented by additional wave gradients during the readout which lead to corkscrew k-space trajectory with 2D CAIPIRINHA sampling scheme. The forward model can be expressed as a generalized SENSE [1] model,

$$wave[x, y, z] = Em[x, y, z] \quad (1)$$

where E is the encoding matrix, specifically,

$$E = MF_x^{-1}Psf[k_x, y, z]F_x S \quad (2)$$

where $F_x$ is Fourier transform in readout direction; S is the sensitivity encoding matrix; $Psf[k_x, y, z]$ is the point spread function (PSF) encoding caused by the wave gradients; M is the aliasing matrix due to 2D CAIPIRINHA undersampling.

Wave-CAIPI fully exploits coil sensitivity variations in three directions. So, as seen as Fig. 1(c), it significantly reduces g-factor penalty and yields the g-factor maps close to unity for highly accelerated 3D imaging.

### 2.2. The G-factor Value in the Arbitrary Position
The theoretical calculation method [11] of the g-factor map has been extended to Wave-CAIPI imaging with the closed-form as [11],

$$g_\rho = \sqrt{(E^H E)_{\rho,\rho}^{-1}(E^H E)_{\rho,\rho}} \quad (3)$$

where E is the encoding matrix; the subscript $\rho$ is the spatial position.

The direct calculation of the Eq.(3) is extremely time-consuming and memory-consuming for the large-scale matrix inverse. To raise computational efficiency and reduce memory-consuming, iterative calculation is used to calculate g-factor values in the arbitrary position. Specially,

$$g_\rho = \sqrt{e_\rho^H d}\sqrt{e_\rho^H(E^H E)e_\rho} \quad (4)$$

where $e_\rho$ is a vector with the element in the $\rho$ position being 1 and others being 0; the vector d is calculated by iteratively solving the linear sub-problem as follows,

$$(E^H E)d = e_c \quad (5)$$

### 2.3. Approximate Calculation of Average G-factor
The average g-factor penalty is evaluated by,

$$g_{mean} = \frac{1}{N_\rho}\sum_{\rho=1}^{N_\rho} g_\rho \quad (6)$$

where $N_\rho$ is the number of voxels in the ROI; $g_\rho$ is the g-factor value in $\rho$ position.

Because Wave-CAIPI acquisitions spread the aliasing evenly and take full advantage of the coil sensitivity variation in three directions, it yields the g-factor maps close to unity [10]. There are low-order polynomial functional relationship between the g-factors in all spatial positions. From the Eq. (6), it can be further known that the average g-factor has low-order polynomial functional relationship with the g-factor in the arbitrary position (e.g. the central position). Here, the g-factor penalty in the central is used. And we assume that this functional relationship is,

$$g_{mean} = f(g_c) \quad (7)$$

The range of $g_{mean}$ and $g_c$ are very small and $f(1) = 1$ since $g_c = 1$, when $g_{mean} = 1$. For instance, $g_{mean} \in [1,2]$ and $g_c \in [1,3]$ in our experiments. Therefore, this function can be expanded by Taylor series [19] at the point (1,1) as,

$$g_{mean} = \frac{f(1)}{0!} + \frac{f'(1)}{1!}(g_c - 1) + \cdots + \frac{f^{(n)}(1)}{n!}(g_c - 1)^n \quad (8)$$

It is well approximated by the first order term, which is called as the Taylor linear approximation [19],

$$g_{mean} \approx 1 + f'(1)(g_c - 1) = 1 + \eta(g_c - 1) \quad (9)$$

where $\eta$ is the coefficient of the first-order term and $g_c$ is rapidly calculated using the Eq. (4) and (5).

According to our current experiments, the value of $\eta$ is insensitive to the parameters of wave gradients (relative amplitudes and cycles). This value of $\eta$ is usually set to 0.3~0.5 empirically in practice. Alternatively, it can also be estimated by a few samples, $(g_c^1, g_{mean}^1), (g_c^2, g_{mean}^2), \ldots, (g_c^k, g_{mean}^k), \ldots, (g_c^n, g_{mean}^n), k = 1, \ldots, n$. In the latter case, while $g_c^k$ can be calculated using the Eq. (5) and (6), $g_{mean}^k$ can be rapidly and approximately calculated by

$$g_{mean}^k = \frac{1}{N_\Omega} \sum_{\rho \in \Omega} g_\rho \quad (10)$$

where $\Omega$ is a set made up of randomly distributed positions in the ROI; $N_\Omega$ is the number of elements in $\Omega$.

## 3. METHOD

Here, we firstly illustrated the simulation experiments to verify the proposed fast calculation method of average g-factor (verification experiments). And then, phantom and in vivo experiments were performed to compare the proposed method with the theoretical calculation method and pseudo multiple replica method in terms of time cost (comparison experiment). In the verification experiments of in vivo human brains, Wave-CAIPI imaging with different wave gradient parameters (relative amplitude and cycles) was simulated. Specifically, the relative amplitude was at the range of $[0.4, 20]$ with the interval of 0.4 and the number of cycles was at the range of $[1, 20]$ with the interval of 1. These simulations were performed on the data acquired by Wave-CAIPI 3D GRE sequence with the following protocol parameters: resolution was $1.0 \times 1.0 \times 1.0 \text{mm}^3$; FOV was $192 \times 192 \times 192 \text{mm}^3$; $TR = 26 \text{ ms}$; $TE = 13 \text{ ms}$; flip angle= $9°$; bandwidth= 50 Hz/pixel. And the acceleration factors were retrospectively $R_y \times R_z = 3 \times 3$. Among the comparison experiments, the three brain data of different volunteers were acquired by the same protocol as above, except that the FOVs were $210 \times 210 \times 120 \text{mm}^3$. The average g-factors were calculated by the theoretical calculation, the pseudo multiple replica and the proposed method respectively. All results were performed in the HP Z820 workstation with the Intel Xeon E5-2640 CPU and 128 GB memory.

## 4. RESULTS

Fig. 2 (a) and (b) show the three-dimensional surface graphic of the average g-factors with varying wave gradient parameters (relative amplitude and cycles). They were calculated by using the pseudo multiple replica method and the proposed method respectively. Their two-dimensional view is shown in Fig. 2 (c). Although the proposed method uses some approximate calculations, the average g-factor penalty is well approximated. As see as Fig 2, the average g-factor calculated by the proposed method is consistent with that calculated by the pseudo multiple replica method. Although the coefficient of the first-order term ($\eta$) in the Taylor linear approximation might be specific to many system elements, it can be set to an empirical value in practice for the current Wave-CAIPI frameworks which yield g-factor maps close to unity [10] and thus the calculation error is very small. Here, $\eta = 0.37$ was used in all experiments. If the relative amplitude of wave gradients is too small, that condition, Wave-CAIPI yielding the g-factor maps close to unit, is not fulfilled. Then, the calculation error is large when using the empirical $\eta$. In that case, $\eta$ should be more precisely calculated by the Eq. (10), as mentioned above.

Table 1 shows the comparisons of the proposed fast calculation method with the theoretical calculation method and the pseudo multiple replica method in terms of time cost. The three experiments illustrate that it is averagely 1678 times much faster than the pseudo multiple replica method, and 994 times faster than the theoretical calculation method.

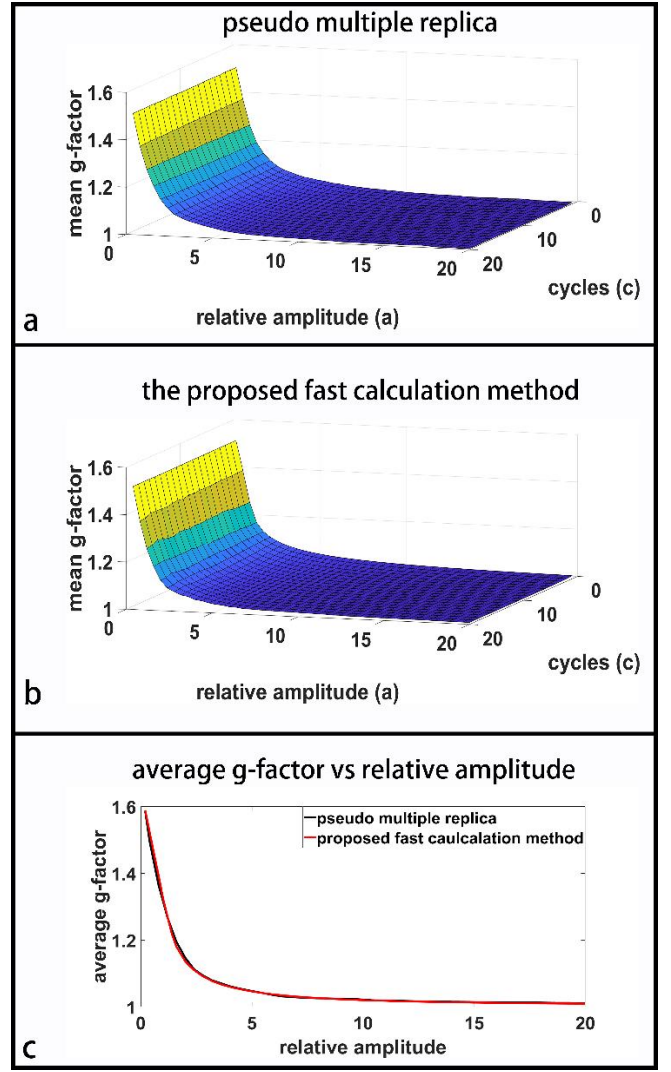

Fig. 2. 3D surfaces of the average g-factors with varying relative amplitude and cycles of wave gradients. They were calculated by using the pseudo multiple replica method (a) and the proposed fast calculation method (b) respectively. Their 2D views are shown for comparison (c).

Table 1. The time cost of the theoretical calculation, pseudo multiple replica and proposed methods

| In vivo human brain experiments | Theoretical calculation method | Pseudo multiple replica | Proposed calculation method |
|---|---|---|---|
| Exp. 1 | 29min41s | 52min23s | 2.23s |
| Exp. 2 | 39min04s | 51min47s | 1.86s |
| Exp. 3 | 19min32s | 41min20s | 1.27s |

## 4. CONCLUSION

In sum, a novel fast calculation method was proposed to compute the average g-factor for Wave-CAIPI MR imaging. The proposed method firstly calculates the g-factor value in the arbitrary (e.g. the central) position separately and then approximate it to the average g-factor using the Taylor linear approximation. The verification experiments illustrated that the proposed method of the average g-factor penalty was much faster than the conventional theoretical calculation and pseudo multiple replica methods. In the future, the proposed method will be applied into further optimizing the application parameters and sampling patterns of Wave-CAIPI MR imaging.